\documentclass[preprint]{aastex631}

\usepackage{multirow}
\accepted{2024 Nov}

\begin{document}

\title{Boulder migration in the Khonsu region of comet 67P/Churyumov-Gerasimenko}

\correspondingauthor{Xian Shi}
\email{shi@shao.ac.cn}

\author[0009-0000-9322-4473]{Xiang Tang}
\affiliation{Shanghai Astronomical Observatory, Chinese Academy of Sciences \\
80 Nandan Road, 200030 Shanghai, China}
\affiliation{School of Astronomy and Space Sciences, University of Chinese Academy of Sciences \\
19A Yuquan Road, Beijing 100049, China}

\author[0000-0002-4120-7361]{Xian Shi}
\affiliation{Shanghai Astronomical Observatory, Chinese Academy of Sciences \\
80 Nandan Road, 200030 Shanghai, China}

\author[0000-0002-8262-0320]{Mohamed Ramy El-Maarry}
\affiliation{Khalifa University of Science and Technology \\
P.O. Box: 127788, Abu Dhabi, UAE}

\begin{abstract}
European Space Agency’s Rosetta mission is the only space mission that performed long-term monitoring of comet at close distances. Its over two years’ rendezvous with comet 67P/Churyumov-Gerasimenko revealed diverse evolutionary processes of the cometary nucleus. One of the most striking events is the migration of a $\sim$30-m boulder in the southern hemisphere region of Khonsu. Previous works found the boulder's $\sim$140-m displacement occurred during the three months from August to October 2015, and several triggering mechanisms were proposed, including outburst at the boulder site, seismic vibrations from nearby activities, or surface erosion of the slope beneath the boulder. In this work, we further analyze this impressive event by analysing imaging data from Rosetta’s OSIRIS camera. We constrained the boulder's migration time to within 14 hours and derived a detailed timeline of the boulder migration event and local dust activities. High-resolution thermophysical modelling shows significant dichotomy in the thermal history of the boulder's southern and northern sides, which could have triggered or facilitated its migration via its own volatile activity.
\end{abstract}
\keywords{ Comet nuclei, Comet surfaces, Comet activity}

\section{Introduction} \label{sec:intro}
European Space Agency's Rosetta mission acquired vast amount of high resolution data from close proximity that reveal the evolution of a comet during its over two-years' operation around comet 67P/Churyumov-Gerasimenko (hereafter 67P) \citep{taylor2017rosetta}. Comet 67P is a Jupiter-family comet that orbits the Sun every $\sim$6.5 years. It has a bi-lobed shape consisting of a $\sim$3 km body and a $\sim$2 km head connected by a narrow neck \citep{2015Sci...347a1044S}. One of the most significant findings by Rosetta is the intricate interplay between the cometary nucleus and its coma. The nucleus, being the source of gas and dust coma, is simultaneously being replenished and reshaped by coma materials  \citep{2015A&A...583A..17T,keller2017seasonal}. The repository of pre- and post-perihelion imaging data have allowed us to investigate changes of the nucleus \citep{barrington_quantifying_2023}, as well as their interrelation with cometary gas and dust activities \citep{hu_seasonal_2017}. 

Nominal dust and gas emission from the nucleus are predominately controlled by the diurnal cycle of solar irradiation as the nucleus rotates with a period of $\sim$12.4 h \citep{2015Natur.525..500D}. It forms the observable fine structures of the dust coma that are shaped by local topography and viewing conditions \citep{2018NatAs...2..562S}. Collimated dust features, or dust ``jets", were identified to originate from various morphological features, such as fractured terrains, scarps, pitted texture and also, smooth areas \citep{vincent2016fractured,shi2016sunset,vincent2015large,fornasier2019linking}. In contrast, spontaneous activities, or ``outbursts", are irregular and intense emanation events that could be triggered by different processes including thermal cracking, rupture of deep volatile reservoir, landslides, etc. as observations and models proposed \citep{2016A&A...593A..76S,pajola2017pristine,agarwal2017evidence,vincent2016summer}. 

Dust activities play an important role in redistributing materials around the nucleus. Dust grains and aggregates lifted might not escape the nucleus' gravity but rendezvous in the near-nucleus coma, with a considerable portion ultimately falling back \citep{2015A&A...583A..17T,keller2017seasonal}. Fine dust, with typical sizes of micro-meter to centimeter, can bury different surface features, especially small boulders and outcrops, while some previously covered surfaces can be exposed by dust removal \citep{hu_seasonal_2017, keller2017seasonal,el2017surface,2020A&A...636A..91C}. Decimeter- to meter-sized blocks released from the nucleus have also been observed in the near-nucleus coma, some of which with trajectories falling back towards the nucleus  \citep{agarwal2017evidence,2022A&A...659A.171P,shi2024diurnal}.

For larger chunks, several meters to tens of meters in size, their relocation usually creates more remarkable surface changes. Cliff collapse combined with outburst are found to predominantly contribute to the formation of talus, as the crumbling wall creates new boulders and debris fields \citep{pajola2017pristine}. A boulder with diameter of $\sim$10 m in the neck region could have dislodged from a nearby cliff and bounced several times across the surface, leaving trails in the smooth terrain \citep{2019EPSC...13..502V}. Considering larger chunks can retain a considerable portion of volatile ices in the process of being redeposited, their migration contribute notably to the mass and volatile cycle of the comet \citep{keller2017seasonal}. However, the driving mechanism and development process of such kind of boulder migration are still poorly understood. Especially, how such events are related to local activity, or what role the gas and dust activities play, need further investigation. 

\begin{figure}[h] 
\centering
\includegraphics[width=0.8\linewidth]{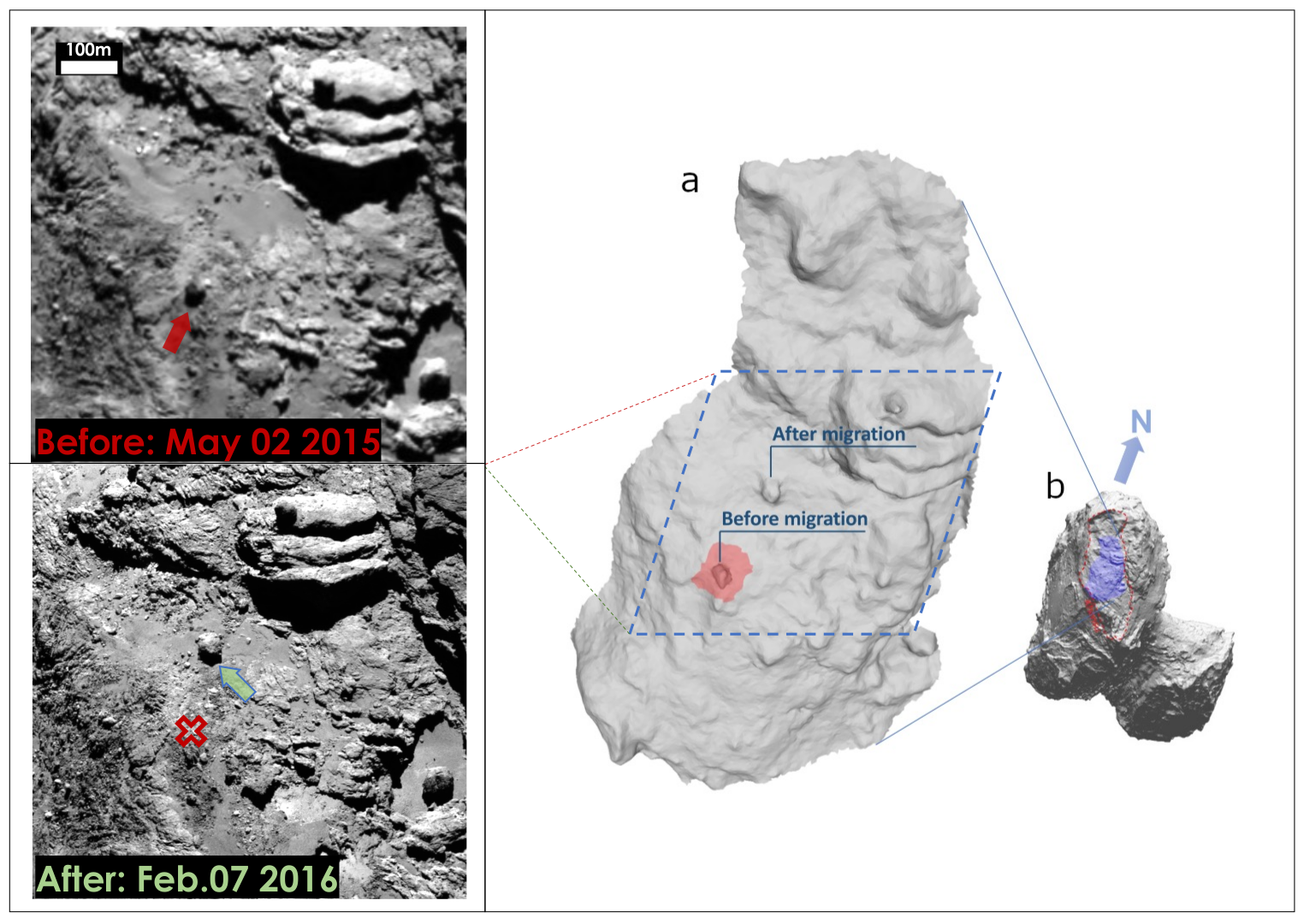}
\caption{High-resolution images and Shape model of the boulder and its surrounding areas. In images, the arrows point to the boulder at that time and the red cross marks the original position of boulder in the `after' image, the dates are the image acquisition time. (a) Local shape model showing the boulder at both locations before and after the migration. Red area covers the localized shape model used for thermal analysis. (b) Location of the Khonsu region (outlined in dashed red line) on the large lobe of the nucleus. Blue area is as shown in decimated shape model a.}
\label{fig:context}
\end{figure}

The arguably most intriguing case of boulder migration is the transport of a $\sim$30-m-diameter boulder by $\sim$140 m, which is by far the largest boulder that is identified with displacement on 67P (Figure \ref{fig:context}, \cite{el2017surface}). The event took place in the Khonsu region of the southern hemisphere of 67P (Figure \ref{fig:context}a, \cite{2018P&SS..164...19T}). Khonsu is one of the most active regions on the nucleus, as numerous outbursts and morphological changes were observed \citep{hasselmann2019pronounced, barrington_quantifying_2023}. In this area the peak of surface changes and activities occurred one month after perihelion (2015 August 13, $r_h = 1.243 $AU)
\citep{hasselmann2019pronounced}. 

The migrated boulder was first reported by \citet{el2017surface}, who narrowed the time of the migration to between 2015 August 1 and October 30. \cite{hasselmann2019pronounced} estimated the activity time to within two months after 67P's perihelion (Figure \ref{fig:joint_plot}).  
The triggering mechanism of such event remains not well understood, while nominal water ice activity should not be sufficient to mobilize such massive boulder. Several hypotheses were proposed to explain the boulder's destabilisation, including: an outburst that directly lifted or pushed the boulder, the erosion of the sloped surface where the boulder resided, or a local seismic vibration from an active source. 
Among them, the most commonly agreed one is that an outgassing from the boulder’s original location or in the neighboring areas, exerted a force on the boulder and triggered its roll downslope \citep{el2017surface,hasselmann2019pronounced}. 

In this work we further investigate this boulder migration event, aiming to determine the time and condition of the movement, together with its potential link to local dust activity. Moreover, by modeling the thermal history of the boulder and its surrounding areas, we discuss possible mechanism for its mobilisation and the implications on cometary material redistribution.

\section{Observations} \label{sec:obs}

We utilise observations acquired by Rosetta's scientific camera system OSIRIS (Optical, Spectroscopic, and Infrared Remote Imaging System)  \citep{keller_osiris_2007}. 
OSIRIS consists of one Narrow Angle Camera (NAC) and one Wide Angle Camera (WAC) with field of view of $2.2^\circ \times 2.2^\circ$ and $11.6^\circ \times 11.6^\circ$, respectively. They acquired image sequences of 67P's nucleus and its surrounding coma with different operation modes fulfilling designated science objectives \citep{2015Sci...347a1044S}. 
We used the Level 3 (CODMAC Level 4) imaging data with same filters and similar exposure times. The images were geometrically and radiometrically calibrated, and archived in PDS format \citep{tubiana_scientific_2015}.

For determining a timeline of the event, we searched through OSIRIS database with the help of ESA's Planetary Science Archive(PSA, \url{https://archives.esac.esa.int/psa/}). A candidate catalog of images was first built by filtering out images that cover the region of interest within the time range of the boulder migration. We then select from the catalog images that display the surface area containing the boulder with favorable viewing geometries. Key images used in this study are listed in Table \ref{tab:image_info}.

 \begin{table}[h]
 \tabletypesize{\scriptsize}
 \tablewidth{0pt} 
 \caption{Information of selected OSIRIS image data for this study.}
 \label{tab:image_info}
 \begin{tabular}{lccccccc}
 \hline
 \hline
 Sequence(s) (UTC) & Sun lat(\textdegree) & Sun lon(\textdegree) &\textit{ }\textit{$r_h$} ($\rm AU$) & $\alpha$(\textdegree) & $d$($\rm km$) & Res.($\rm m~px^{-1}$) \\ \hline
 NAC 2015-03-25, 19:00 & 15.0 & 232.6 & 2.01 & 72.8 & 83.0 & 1.544 \\
 NAC 2015-05-02, 15:09 & 3.0 & 208.8 & 1.73 & 60.5 & 124.9 & 2.323 \\
 NAC 2015-05-23, 21:51 & -5.7 & 184.9 & 1.57 & 61.8 & 159.3 & 2.963 \\
 NAC 2015-09-26, 06:00 & -48.9 & 328.6 & 1.36 & 53.4 & 771.3 & 14.347 \\
 NAC 2015-10-03, 09:26 & -46.7 & 303.6 & 1.39 & 52.3 & 1072.9 & 19.956 \\
 NAC 2015-10-03, 19:30 & -46.6 & 6.4 & 1.39 & 52.7 & 1021.1 & 18.992 \\
 NAC 2015-10-03, 22:49 & -46.6 & 268.0 & 1.39 & 52.9 & 1004.1 & 18.676 \\
 NAC 2015-10-04, 19:01 & -46.3 & 30.9 & 1.40 & 54.0 & 903.0 & 16.797 \\
 NAC 2015-10-04, 19:21 & -46.3 & 20.9 & 1.40 & 54.0 & 901.4 & 16.766 \\
 NAC 2015-10-04, 19:41 & -46.3 & 11.1 & 1.40 & 54.0 & 899.8 & 16.736 \\
 NAC 2015-10-04, 20:01 & -46.3 & 1.3 & 1.40 & 54.1 & 898.1 & 16.705 \\
 NAC 2015-10-04, 20:21 & -46.3 & 351.3 & 1.40 & 54.1 & 896.5 & 16.675 \\
 NAC 2015-10-04, 21:18 & -46.2 & 323.4 & 1.40 & 54.2 & 891.9 & 16.589 \\
 NAC 2015-10-09, 09:39 & -44.7 & 357.3 & 1.42 & 59.9 & 580.2 & 10.792\\
 NAC 2015-12-13, 01:14 & -22.0 & 334.1 & 1.87 & 89.3 & 101.0 & 1.879 \\
 NAC 2016-02-07, 13:10 & -7.8 & 220.9 & 2.31 & 59.2 & 47.0 & 0.875 \\
 NAC 2016-02-10, 13:23 & -7.2 & 226.3 & 2.33 & 65.1 & 49.8 & 0.925 \\
 NAC 2016-08-15, 11:19 & 16.86 & 228.9 & 3.58 & 89.0 & 8.9 & 0.166 \\
 \hline
 \end{tabular}
 \tablecomments{Column description: acquisition date of the NAC images, sub-solar point latitude and longitude, 67P heliocentric distance ($r_h$), phase angle ($\alpha$), spacecraft-comet distance ($d$) and surface resolution.}
 \end{table}

\begin{figure}[h]
\centering
\includegraphics[width=\textwidth]{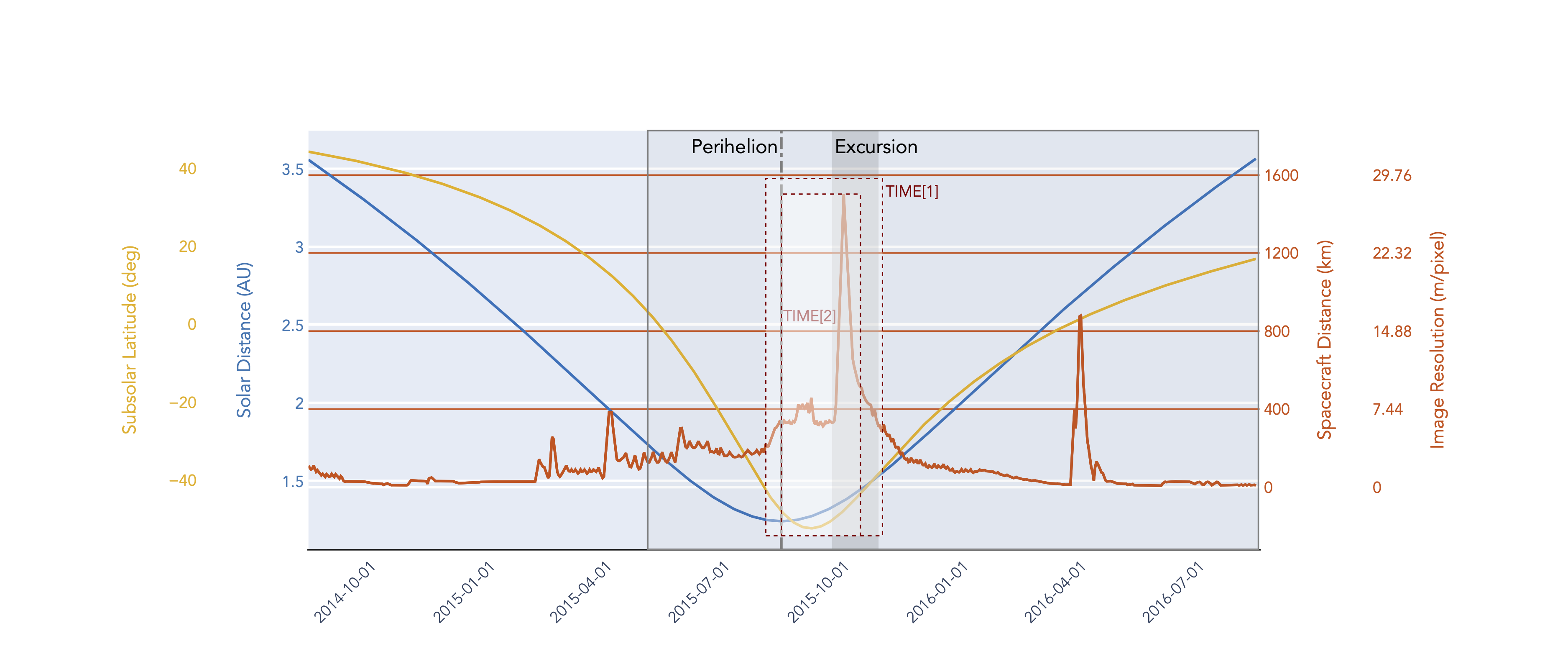} 
\caption{Variations of sub-solar point latitude (degree), 67P's heliocentric distance (AU), spacecraft-comet distance (km) and NAC image spatial resolution ($\rm m~pixel^{-1}$) for the time period between September 2014 and August 2016. The grey-outlined box indicates the time span of data used in this study. The red-dashed boxes with time annotations, \textsc{Time[1]} and \textsc{Time[2]}, represent the proposed migration time for the boulder in \citet{el2017surface} and \cite{hasselmann2019pronounced}, respectively. The perihelion passage and excursion period are present in vertical line.}
\label{fig:joint_plot}
\end{figure}

\begin{figure}[h]
\centering
\includegraphics[width=0.8\linewidth]{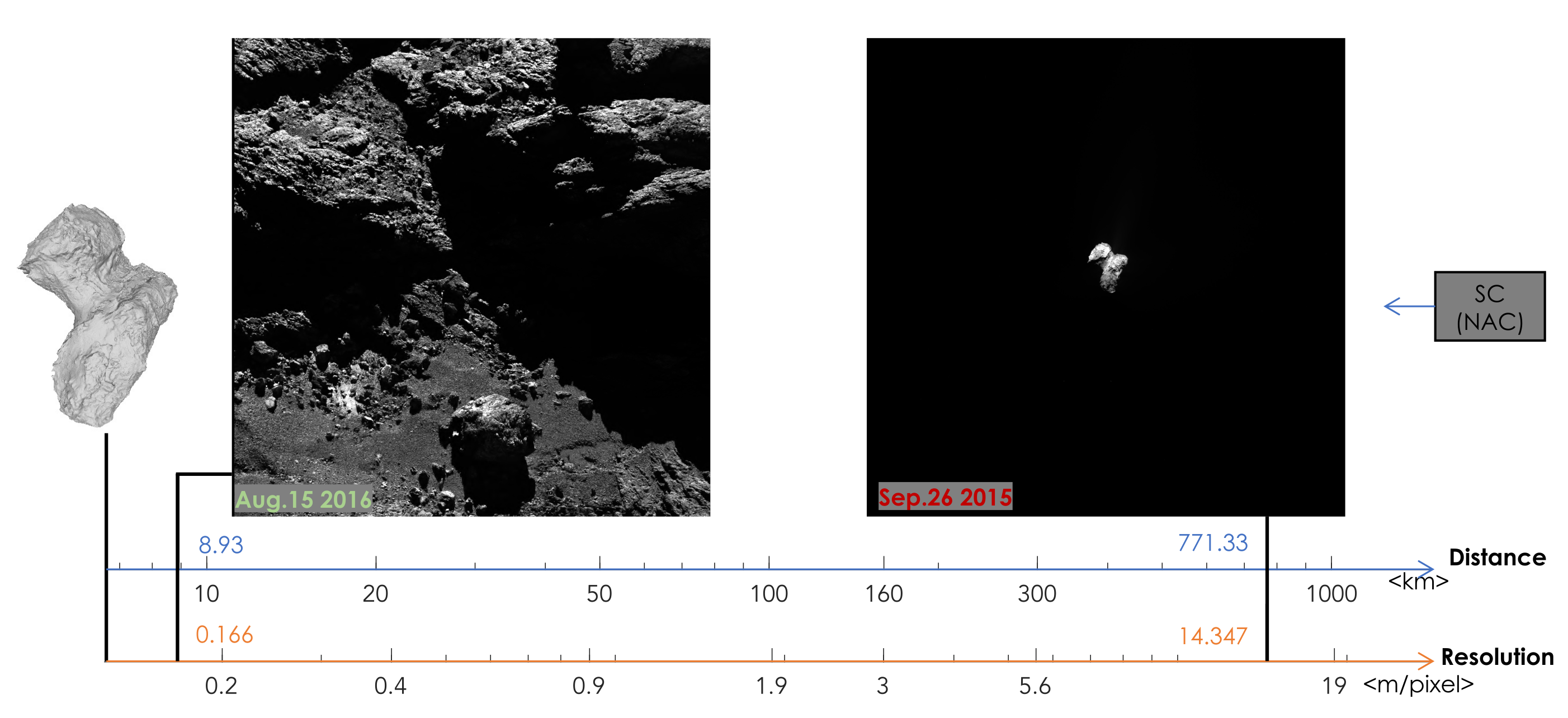} %
\caption{Comparison of NAC surface resolution with Rosetta at different distances to the nucleus. NAC captured two images in different time. The image in rich detail was taken after perihelion while another showing the spacecraft was traveling away from 67P. The axes are set in logarithm scale.}
\label{fig:image_res}
\end{figure}

It is worth mentioning that, OSIRIS images taken during the time of interest are of significantly varying resolutions. At the beginning of 2015 October, Rosetta performed an excursion to the comet tail when the spacecraft reached as far as over 1500 km away from the nucleus (Figure \ref{fig:joint_plot}). As illustrated in Figure \ref{fig:image_res}, NAC image taken on 2015 September 26  at a distance of $\sim$771 km yields a surface resolution of $\sim$14.347 $\rm m~px^{-1}$, compared to $\sim$0.166 $\rm m~px^{-1}$ of that image taken on 2016 August 15. During Rosetta's extended mission after perihelion, the spacecraft rendezvoused with the comet at closer distances, enabling us to investigate comet surface features in detail, including the boulder and its surrounding environment after the migration event.

\section{Morphological Analysis} \label{sec:met}

To compare observations acquired under different imaging conditions, and to perform quantitative topographic and photometric analysis, we utilised the stereo-photogrammetric (SPG) SHAP7 shape model of 67P  \citep{preusker_global_2017}. Specifically, we use the shape model version comprising 1 million vertices and 2 million triangular facets, corresponding to a horizontal resolution of $\sim$4 meter. The SHAP7 shape model was derived from images acquired between August 2014 and February 2016, covering almost the entire mission time of Rosetta at comet 67P. As a result, it presents dual existence of the migrating boulder at both its `before' and `after' locations (Figure \ref{fig:context}a). This allows us to identify the boulder's position by matching synthetic images with actual images, compensating for the low resolution of images taken during the excursion. Synthetic images were produced by simulating the view of the camera according to the observational geometry at the time when actual images were captured. Basically, two types of synthetic images were produced: the first with artificial lighting, where all surface features are visible, even those actually in shadow; the second with realistic illumination condition by applying ray-tracing technique. The second type was directly compared with actual images to identify any difference concerning the boulder, and the first type was used for locating the boulder, or other features of interest when they are in the shadow.

We first visually inspected candidate images with pre- and post-migration views of the area to identify possible movement of the boulder relative to common surface features. We then performed geometric calculation to infer the boulder's coordinates on 3D shape model using the position and pointing parameters of the camera provided by the SPICE library \citep{acton2016spice}. In the same way, we also identified other surface changes as well as local dust activities. 

\subsection{Time of the migration} \label{subsec:boulder} 

\begin{figure}[h]
\centering
\includegraphics[width=0.8\linewidth]{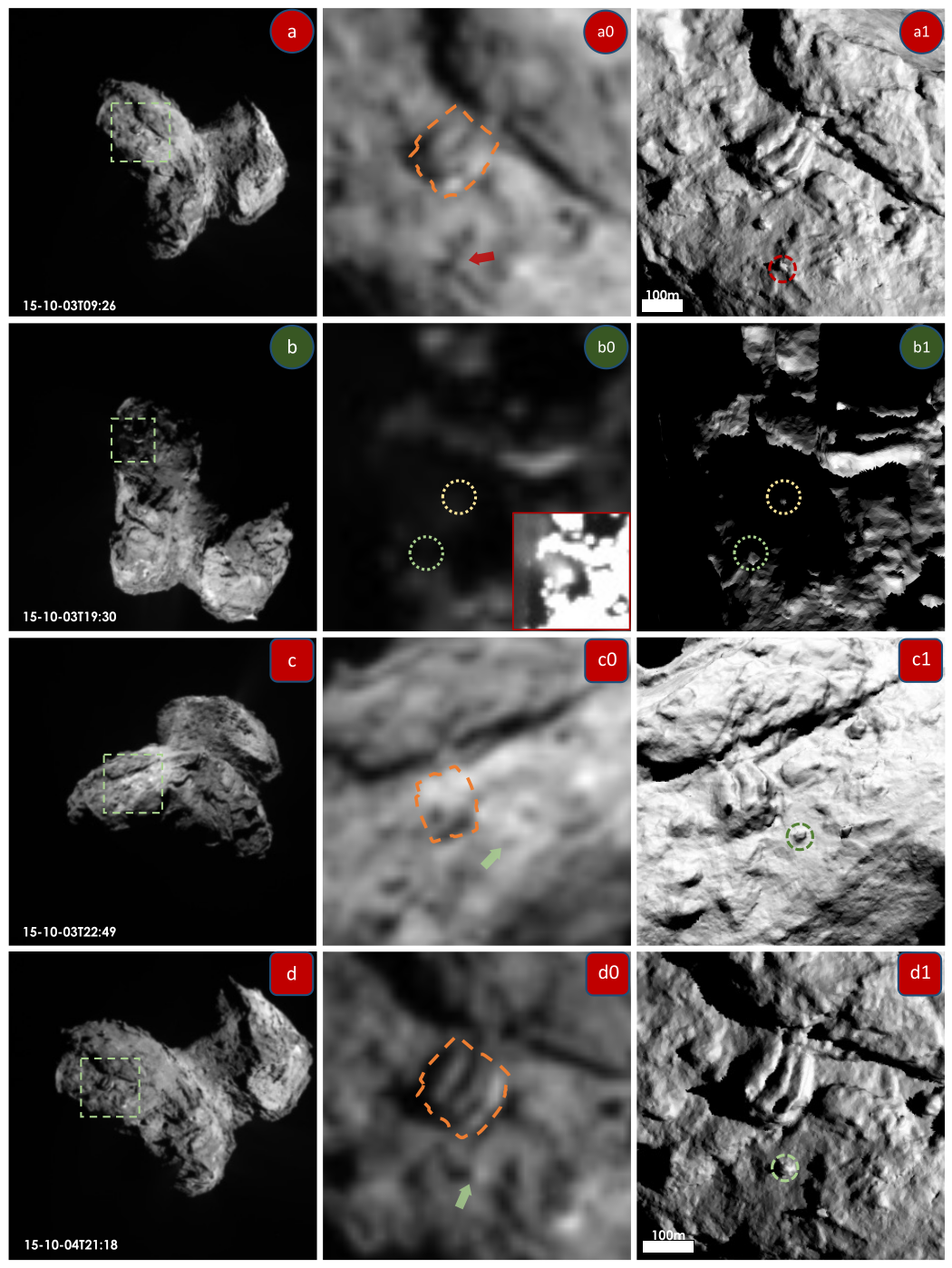}
\caption{OSIRIS observations of the boulder location in early-October of 2015. During this time period images capturing from far distances, we utilized the zoom-in views (second column) and synthetic views (right column) of the observations for comparison. Images with round labels are observations of the boulder before its migration, in which (a) the last image showing that the boulder at its original position. (b) The only image captured during boulder’s migration (with green label).  The circles mark the position of boulder before and after its migration. The position of the boulder cannot be identified  as a result of unclear lighting condition. For comparison, in the bottom right in b0 is the contrast-stretched view of the actual image. The square image labels represent post-migration observations of the boulder, where (c) fairly low-resolution showing the boulder at its new position, (d) an image of the first confirmation for the migrated boulder. Red arrows and red hollow cross mark the original location of the boulder, green arrows point to the boulder after its migration and orange dash lines identify the ``pancake" feature.}
\label{fig:boulder_migrate}
\end{figure}

We discovered that the migration occurred on 2015 October 3, roughly fifty days after 67P’s perihelion point. Since the Rosetta spacecraft was at a distance over one thousand kilometers from the nucleus, images taken during that time have typical resolutions of $\sim$ 20 $\rm m~px^{-1}$ (Table \ref{tab:image_info}), in which the boulder extends over only two pixels. However, the presence of boulder shadows on the surface helped us identifying the boulder \citep{oklay2016comparative}. The two NAC images taken at UTC 09:26 and 22:49 (Figure \ref{fig:boulder_migrate} a0 \& c0) presents the comparative view closest in time before and after the boulder's migration. By blinking between real and synthetic images, we identify the boulder was at its original location at 09:26 with its lengthened shadow (Figure \ref{fig:boulder_migrate} a0 \& a1), while at 22:49, the boulder had disappeared from its initial location, but leaving faint shadow at it new position (Figure \ref{fig:boulder_migrate} c0 \& c1). This constrains the migration time to within 14 hours, slightly longer than one rotation, indicating that the migration event was not due to long-term evolution of the boulder, but rather a sudden and rapid displacement likely related to local activity.  

Only one other OSIRIS image  covering this area was acquired in between, at 19:30 on October 3. Unfortunately, the region was at night at that time and surface features are poorly illuminated (Figure \ref{fig:boulder_migrate}b). The observational conditions in original image cannot directly provide detect any clues for the migration event.  Though this image actually recorded information of local dust activities (see subview in Figure 4b, and Sec. \ref{subsec:activity}). Another image acquired later at 21:18 on October 4 shows the area under favourable illumination (Figure \ref{fig:boulder_migrate}d), and the boulder could be confirmed to be unmistakably at its new location. 

\subsection{Boulder and its surroundings} \label{subsec:geomorphology}
\begin{figure}[h]
\centering
\includegraphics[width=\textwidth]{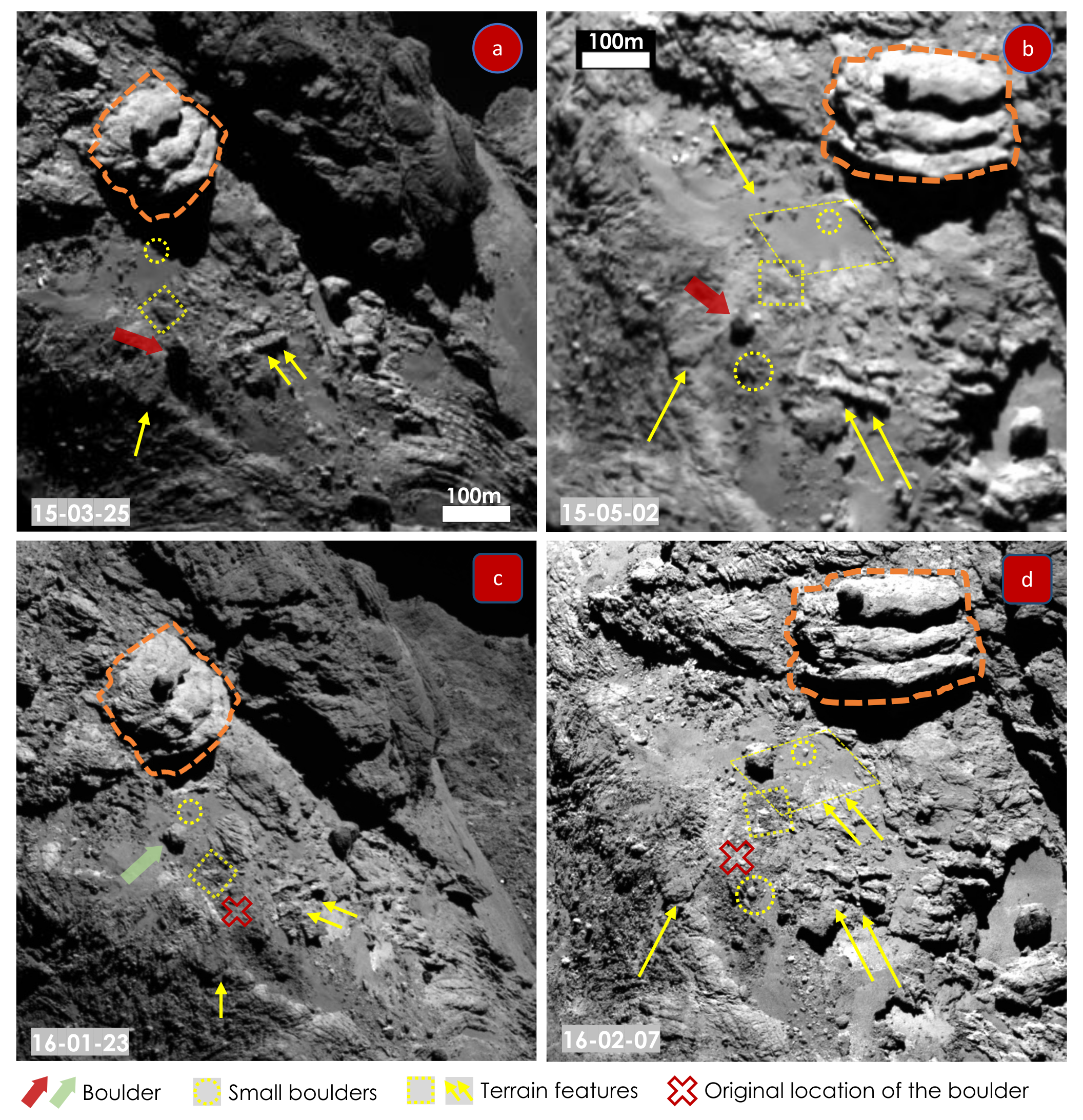}
\caption{OSIRIS observations of the boulder and its surroundings. Yellow arrows highlight the cluster of small boulder, the shallow scarp, the cluster of outcrop, situated to the north, west and east of the boulder’s former location, respectively. Yellow circles mark the presence of the isolated small boulder, and rectangles mark a lumpy relief in the possible route of migration. (a) side view observation of Khonsu region with the best spatial resolution before perihelion. (b) front view of boulder’s neighbouring areas. (c) observation of post-perihelion morphological state of local region from side view. (d) front view for inspecting the boulder's surrounding after perihelion. Red arrows and red hollow cross mark the original location of the boulder, green arrows point to the boulder after its migration and orange dash lines identify the ``pancake" feature.}
\label{fig:env_evolve}
\end{figure}

We compare pairs of images with similar observation geometry acquired before and after the migration event to investigate changes in the morphology and positions of the boulder and its surroundings (Figure \ref{fig:env_evolve}). Numerous changes were detected including mass excavation processes and cases of instability, conforming with conclusions of previous studies that Khonsu is one of the most active zones on the nucleus \citep{hasselmann2019pronounced,2019A&A...630A...7F}.
Before its migration, the boulder situated between a shallow scarp and a cluster of outcrops (highlighted in Figure \ref{fig:env_evolve}b \& d), on a hill with rough texture. To quantitatively evaluate the boulder's original dynamic environment, we chose 370 facets on the slope between the boulder's original and final positions \ref{fig:ks_slp}. The average effective slope of these facets is $\sim$25.5$^\circ$ with a typical value of $\sim$23$^\circ$ (Figure \ref{fig:ks_slp}b), compared to the global typical value of $\sim$14$^\circ$ over the 2 million facets. Close to the original position of boulder, the effective surface slope can reach $\sim$46$^\circ$. Therefore, the dynamic topography around the boulder could facilitate its motion once it was destabilized. After the migration, the boulder ended up in a smooth terrain near the ``pancake" feature, surrounded by a cluster of small boulders. 
This smooth terrain covered $\sim$ 135 $\times$ 100   m$^{2}$ before perihelion, in which a small boulder was almost buried (Figure \ref{fig:env_evolve}b). After perihelion, however, this area shrunk as the top unconsolidated layer was likely exhumed, revealing the complex features on its outskirts along with the exposure of the small boulder (Figure \ref{fig:env_evolve}b \& d, marked by small circles near the ``pancake" feature). 

Unlike the bouncing boulder in the comet's smooth neck region \citep{2019EPSC...13..502V}, as the migrating boulder moved northward downslope (Figure \ref{fig:ks_slp}), its possible path displayed no clear track other than some mild mass removal. Only a lumpy relief located within the surface slope was detected have been eroded (Figure \ref{fig:env_evolve}a \& c, marked by small rectangles). Erosion processes were quite common in the nearby region. To the east of the boulder’s original location, the cluster of outcrops broke into three segments, accompanied with the exposure of several bright patches (Figure \ref{fig:env_evolve}d). A small boulder on the south of the slope where the boulder originally stood appeared to have reoriented due to certain perturbation (Figure \ref{fig:env_evolve}b \& d, marked by small circles next to the original location of the migrating boulder). The lack of track could have different indications depending on the migration scenario of the boulder. It would be a natural consequence if the boulder was ejected from its original location into the near-nucleus space and subsequently fell back on the nucleus at its new location. In the case that the boulder was triggered and rolled downslope, the absence of obvious trail on the ground could imply a relatively high compressive strength of the surface. However, the observation of `no clear track' is limited by the resolutions of pre-migration images, that are lower than 1.5 $\rm m~pixel^{-1}$. If a clear mark was left, like those left by the Philae lander on its way to the final landing site \citep{orourke_philae_2020}, it would serve as perfect experiments to derive the mechanical properties of the cometary material in the local surface. 

\subsection{Local dust activity} \label{subsec:activity}
\begin{figure}[h]
\centering
\includegraphics[width=\textwidth]{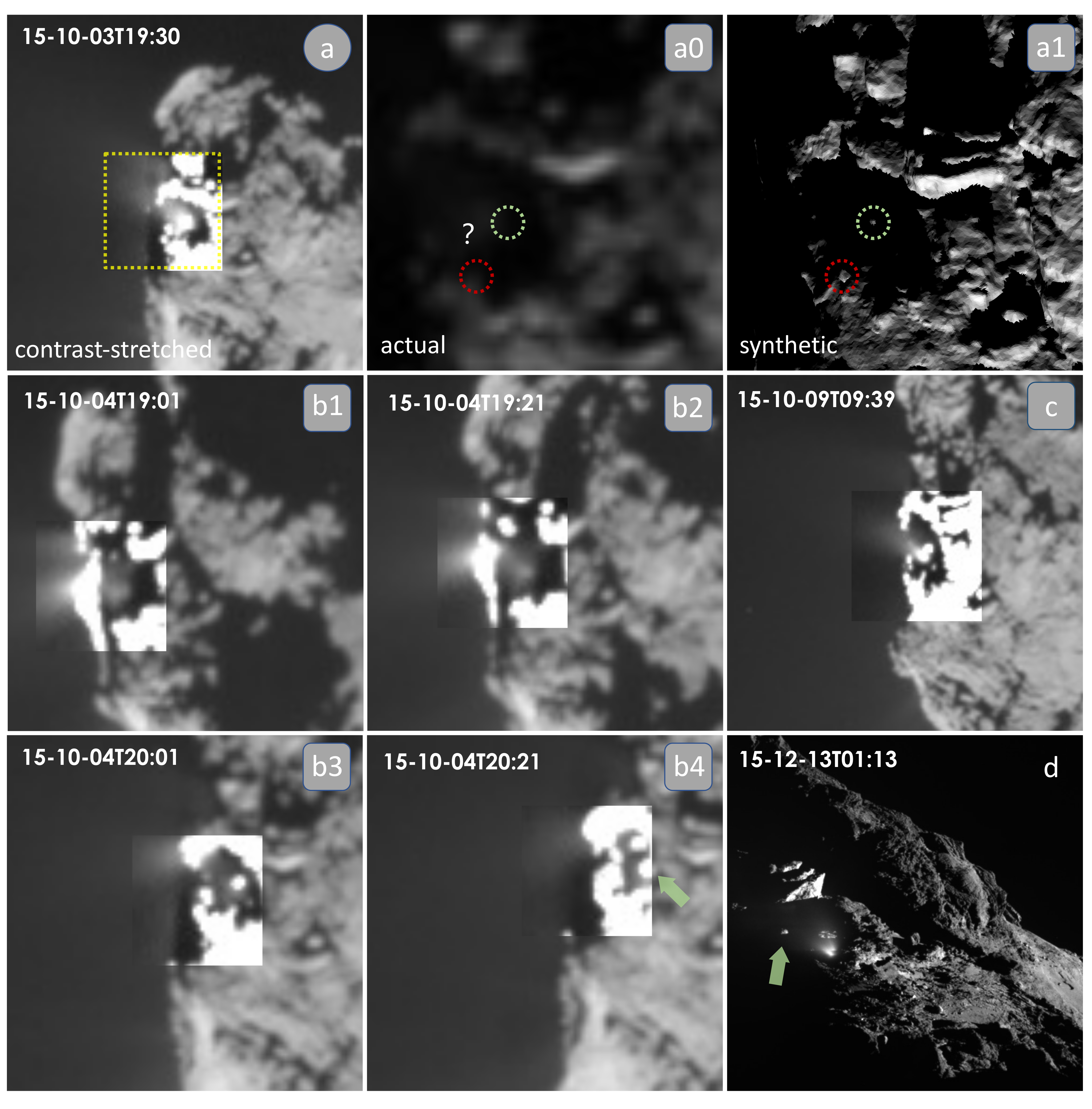}
\caption{Activities related to the migrating boulder in Khonsu region in early and after 2015 October. The cometary activities parts in the images are substituted with the contrast-stretched counterparts for better visualization. (a) the essential and only image for clues of missing boulder in the short time frame when the boulder made its moving, $a$ is the brightness-enhanced image for identifying the dust emission. The activity event is indicated with dash lines. $a0$ is the original image and $a1$ is the synthetic version of local observation. (b) the sequence captured right next two comet rotations after the disappearance of the boulder, showing a prolonged activity after migration event as well as the indication of the location of boulder. Only four images from the sequence are present in this mosaic. (c) an apparent jet next to the boulder and (d) a well-investigated outburst recorded in 2015 December 13 (\cite{hasselmann2019pronounced,el2017surface}, Supp. Mat. S5). Most dust activities appeared in night condition, making dust activities in this restricted area really special. Ejected dust particles dispersed from the badly illuminated area and thus presented an increase in brightness in images.}
\label{fig:act_timeline}
\end{figure}

Our inspection through the imaging data also reveal that the Khonsu region was a site of frequent night-time dust activity. In the image taken at UTC 19:30 on 2015 October 3 (Figure \ref{fig:boulder_migrate}b), a mini-outburst is detected emanating from the smooth terrain near the ``pancake" feature (Figure \ref{fig:act_timeline}a), when we stretched the contrast of the image. The mini-outburst showed a disperse morphology with the dust particles spreading out in all directions from a center point. The brightness in the center of dust emanation was at the same order as the  general surface, one order higher than the dark environment. We constrained its source location by intersecting the boresight of the camera with the shape model. The inferred emission site is near the boulder’s new location (see projected image in Figure \ref{fig:mark_oct}), similar to the condition of the ``mini-outburst" found in Imhotep region early in the mission \citep{knollenberg_mini_2016}. However, due to the low resolution of $\sim$19 $\rm m~px^{-1}$ and the uncertainty in spacecraft orientations, it is unclear whether the outburst is from the unilluminated boulder itself or its surroundings. Here the boulder had already disappeared from its original location as inferred from the comparison between actual and synthetic images (Figure \ref{fig:act_timeline}a0 \& a1). Whether it reached its new location is debatable. According to the synthetic image, top of the boulder should be illuminated if it was already at the new location, which is not the case in the actual image. However, the identification of the boulder at its new location is subject to the accuracy of the shape model. If the boulder’s actual dimensions are smaller than that represented in the shape model, its appearance in the synthetic image would not necessarily indicate its the new location. Since the migration path was not illuminated, it is unclear whether the boulder was in motion or had already reached its new location. 

Roughly two rotations later, on 2015 October 4, a sequence of images with cadence of 10 minutes show the development of another mini-outburst in the same area close to the new location of the boulder (four frames are shown in Figure \ref{fig:act_timeline} b1 to b4, the sequence in appendix). Again, the smooth terrain containing the newly located boulder and the source of the mini-outburst was in darkness during the time, and we could only identify from synthetic images that the source of the activity is in immediate proximity to the boulder or on it (Figure \ref{fig:minioutburst}).

\begin{figure}[h]
\centering
\includegraphics[width=0.9\linewidth]{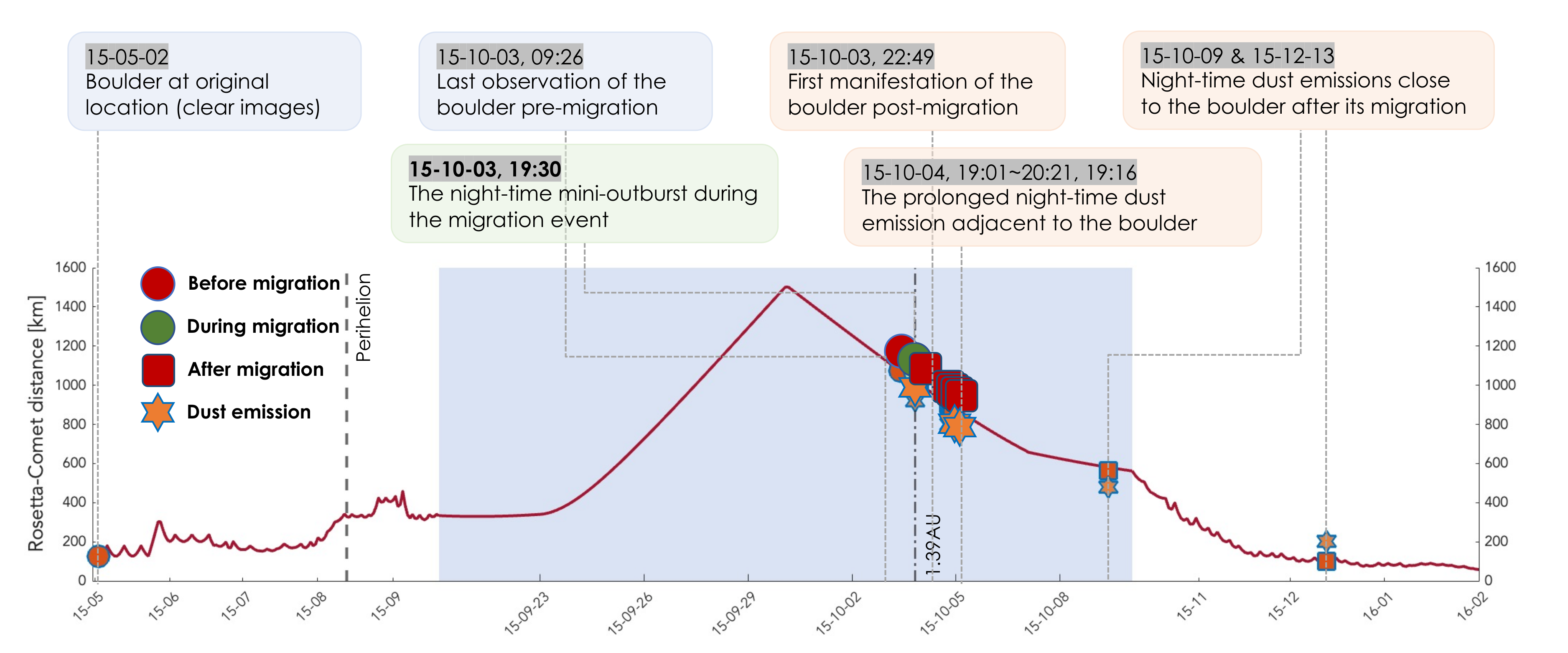} 
\caption{Timeline combining the boulder migration event and local dust activities with their spatial and temporal correlations. The events are marked on comet-spacecraft distance variation. The blue background displays the zoom-in of timestamps from September 20 to October 10, in which local surface changes and activities peaked after comet's perihelion.}
\label{fig:timeline}
\end{figure}

Roughly six days after the boulder migration, on 2015 October 9, a jet appeared next to the illuminated boulder (Figure \ref{fig:act_timeline}c), which was more probably in the peak stage of dust ejection phenomena similar to mini-outburst documented by \cite{knollenberg_mini_2016}. Later on December 13, in similar insolation condition  a massive night-time outburst originated from the patch on the edge of the cluster of lineaments near the original position of the boulder (Figure \ref{fig:act_timeline}d, \cite{hasselmann2019pronounced,el2017surface}). 
A detailed timeline of events, including the evolution of local environment of migrating boulder as well as the dust activities is summarized in Figure \ref{fig:timeline}  .

\section{Thermophysical analysis} \label{sec:ana_thm}

Given the better-constrained migration time of the boulder, we are able to investigate its thermal condition leading to the event. We performed facet-wise thermophysical modeling with a generic 1-D thermophysical code \citep{hu_seasonal_2017,hu_thermal_2017,hu_thermophysical_2021}. The Energy input on a certain facet at any epoch is calculated as the instant insolation, which takes into consideration the orientation of the facet, bolometric bond albedo, as well as topographic shadowing effect \citep{hu_thermal_2017}. 

The model is constructed such that both energy and volatile mass are conserved, which allows us to estimate the temporal variation of temperatures as well as ice content at different depths of each facet \citep{hu_thermophysical_2021}. The conservation equations are as follows:
\begin{equation}
    (c_d\rho_d+c_s\rho_s+c_g \rho_g)\frac{\partial T}{\partial t}+c_g Z_g\frac{\partial T}{\partial x}=-\frac{\partial q}{\partial x}-l_g \zeta_g
\label{eq:energy_cons}
\end{equation}
\begin{equation}
    \frac{\partial \rho_g}{\partial t}+\frac{\partial Z_g}{\partial x}=\zeta_g
    \label{eq:mass_cons}
\end{equation}
where $c_d\rho_d$, $c_s\rho_s$, and $c_g \rho_g$ are the products of heat capacity and mass density of refractory dust, water ice, and water vapour, respectively. $Z_g$ is the mass flux of water vapour diffused through the porous top layer of the nucleus. $q$ is the heat flux by conduction, $l_g$ is the latent heat of water vapour, and $\zeta_g$ is the flux of water vapour sublimation or deposition. 
Thermophysical states of the boulder and its surrounding area are investigated on both diurnal and orbital time scale by converging over one rotation or one orbit of the nucleus, respectively. Primary model parameters are listed in Table \ref{tab:thermal_parameter}. Detailed numerical treatment and model parameters could be found in \cite{hu_thermal_2017} and \cite{hu_thermophysical_2021}.

We use a cutout from the SHAP7 shape model of 500 facets containing the boulder at its original location (red-shaded facets in Figure \ref{fig:context}) for the thermal modelling. The meter-scale resolution of the shape model allows the investigation of differential thermal development over the boulder and its surroundings. 

\begin{table}[h]  
\caption{Parameters of the thermophysical model} 
\centering 
\begin{tabular}{l c c c} 
\hline\hline   
 \multirow{2}{5em}{Parameter} & \multirow{2}{5em}{Symbol} & \multicolumn{2}{c}{Value}  
\\ [0.5ex]  
 & & Diurnal & Orbital
\\ [0.5ex]  
\hline   
 Dust heat capacity & $c_d$ & \multicolumn{2}{c}{$1200 \text{ J}\text{ K}^{-1}\text{ kg}^{-1}$}  \\ 
 Water-ice heat capacity & $c_s$ & \multicolumn{2}{c}{$90+7.49 T \text{ J}\text{ K}^{-1}\text{ kg}^{-1}$}  \\ 
 Water vapour heat capacity & $c_s$ & \multicolumn{2}{c}{$1400 \text{ J}\text{ K}^{-1}\text{ kg}^{-1}$}  \\ 
 Compact dust density & $\rho_d$ & \multicolumn{2}{c}{$2000 \text{ kg}\text{ m}^{-3}$}  \\
 Compact water-ice density & $\rho_s$ & \multicolumn{2}{c}{$1000 \text{ kg}\text{ m}^{-3}$}  \\
 Dust volume-filling fraction & $f_0$ & \multicolumn{2}{c}{$0.2$}  \\
 Initial ice volume fraction & $f_0$ & \multicolumn{2}{c}{$0.1$}  \\
 Water latent heat & $l_g$ & \multicolumn{2}{c}{$2.6\times10^{6} \text{ J}\text{ kg}^{-1}$}  \\  
 Step in depth & $\Delta x$ & \multicolumn{2}{c}{$1\text{ mm}$}  \\
 Step in time  & $\Delta t$ & ${t_{\text{P}}}/40000$ & $ {t_{\text{P}}}/2000$ \\
 Nucleus spin period & $t_{\text{P}}$ & \multicolumn{2}{c}{$44655\text{ s}$} \\
\hline 
\end{tabular}  
\label{tab:thermal_parameter}
\end{table}  

\begin{figure}[h]
\centering
\includegraphics[width=\textwidth]{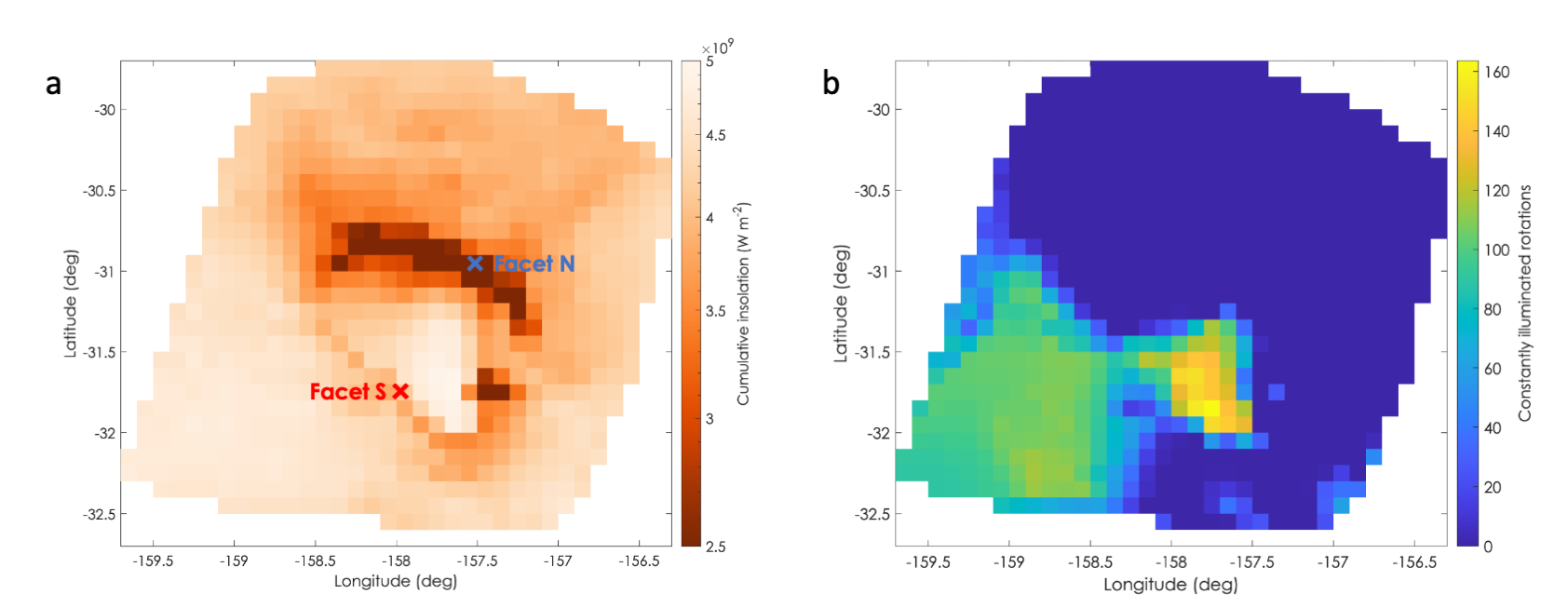}
\caption{Insolation history of the boulder and its surrounding areas before its migration. (a) cumulative insolation between 2014-05-17 and 2015-10-03; The locations of representative facets S and N are indicated with red and blue crosses, respectively; (b) number of rotations with constant illumination, or `polar days', for each facet.}
\label{fig:illumination_map}
\end{figure}

Thermal modelling results show significant dichotomy between the northern and southern sides of the boulder, which is similar to the global dichotomy between the northern and southern hemispheres of 67P due to its large obliquity of $\sim40^{\circ}$. The contrast is essentially induced by received insolation energy. Figure \ref{fig:illumination_map}a shows the map projection of cumulative insolation calculated for the boulder and its surroundings during the five months from May 2014, when 67P was at a perihelion distance of $\sim$4 au, until the occurrence of the migration on at the beginning of October 2015. The average energy received by southern side of the boulder is as much as double that of the northern side. Intriguingly, the southern side of the boulder was subject to constant illumination before the migration event, with some of the facets experiencing constant illumination over more than 150 rotations (Figure \ref{fig:illumination_map}b).

\begin{figure}[h]
\centering
\includegraphics[width=\textwidth]{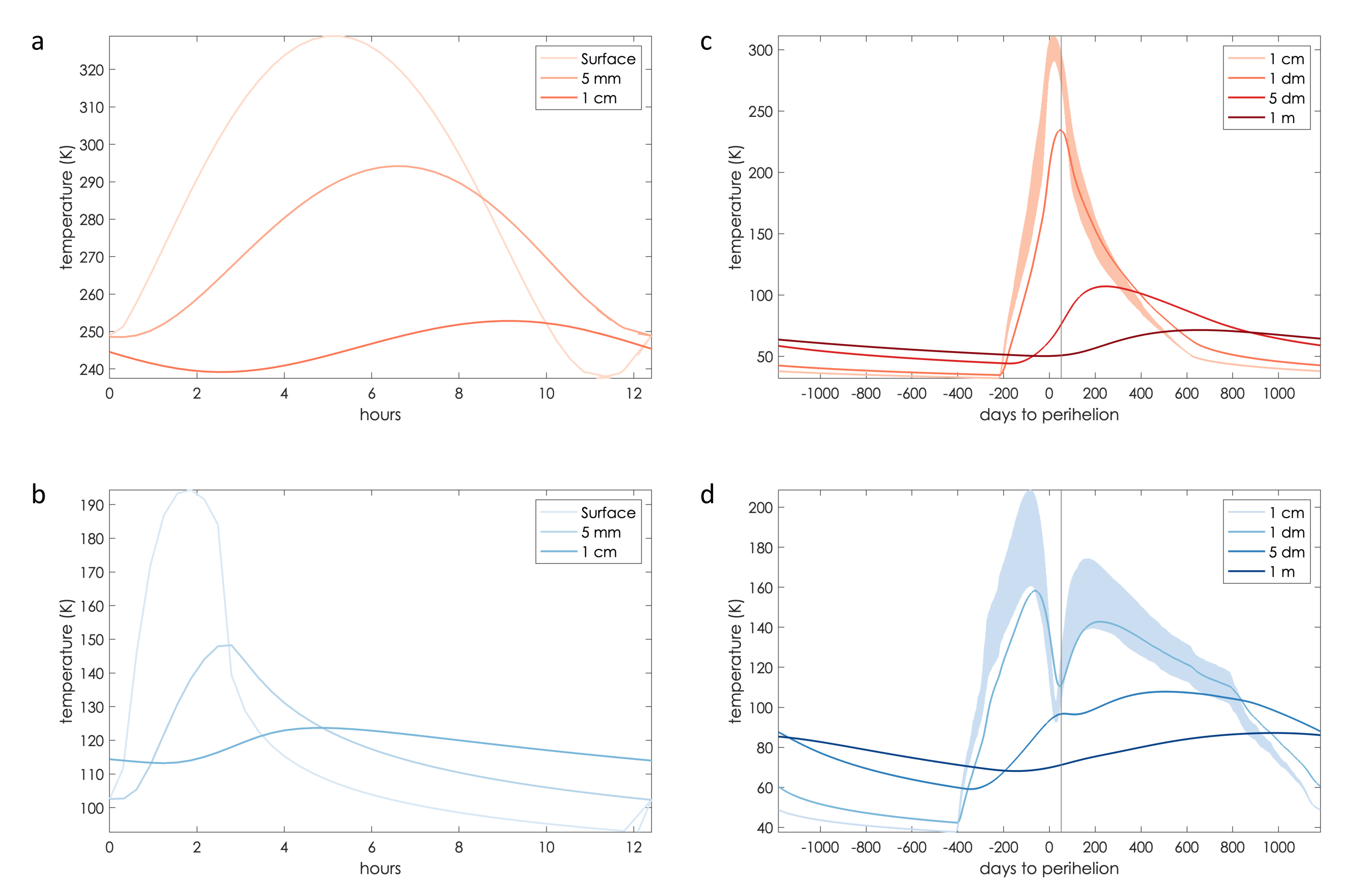}
\caption{Depth-dependent temperature variation at two facets on the southern and northern sides of the boulder during one rotation and one orbit of 67P. (a) Diurnal temperature variation at different depths of facet S from 2015-10-03T10:00:00; (b) the same as (a) for facet N; (c) Orbital temperature variation at different depths of facet S; (d) the same as (c) but for facet N. Vertical gray line indicates the time of the boulder migration.}
\label{fig:thermal}
\end{figure}

For quantitative comparison, we chose two representative facets, facet S on the southern side of the boulder and facet N on the northern side, and modelled the thermal evolution of the top meter of the nucleus at these two facets over both one rotation and one orbit. The diurnal thermal variation was simulated starting at Oct 3 UTC 10:00 to cover the possible migration time of the boulder. Temperatures in the top 1 cm of facet S stayed above 230 K throughout one nucleus rotation, while reaching a maximum of $\sim330 \text{ K}$ (Figure \ref{fig:thermal}a) at $\sim$ UTC 15:00 on October 3. The diurnal variation of temperature gradually attenuated with increasing depth. At $\sim1 \text{ cm}$ depth, roughly where the diurnal heat wave can penetrate, the temperature undulates around $246 \pm 5\text{ K}$, which is still significantly higher than the general sublimation temperature of water ice. On the contrary, facet N on the north face experienced relatively short daytime and received much less energy, resulting in lower temperatures at all depths (Figure \ref{fig:thermal}b). The surface temperature did not exceed $200 \text{ K}$ over the entire rotation, while staying below $170 \text{ K}$ for 70\% of the time. Note that the surface temperature of facet N peaks at an earlier time than facet S, roughly around UTC 12:00. 

On orbital time scale, the dichotomy is even more evident. Figure \ref{fig:thermal} c\&d show variation of temperatures at depths from one centimeter to one meter at the two facets over one orbital period of 6.45 years. Facet S starts to be illuminated from roughly 200 days before perihelion at the end of January 2015. Amount of insolation on the facet increased till around perihelion, then gradually decreased. At the time of the migration event, temperature at one decimeter depth of facet S reached its orbital maximum of 225 K. Conversely, temperature variation in the top one decimeter of facet N shows two peaks, one at $\sim$80 days before perihelion at the end of May 2015, the other at $\sim$150 days after perihelion at the beginning of January 2016. At the time of boulder migration, facet N was experiencing minimum temperatures between the two peaks. The temperature was as low as 110 K at one decimeter deep, reaching its minimum during the perihelion period.

\begin{figure}[h]
\centering
\includegraphics[width=\textwidth]{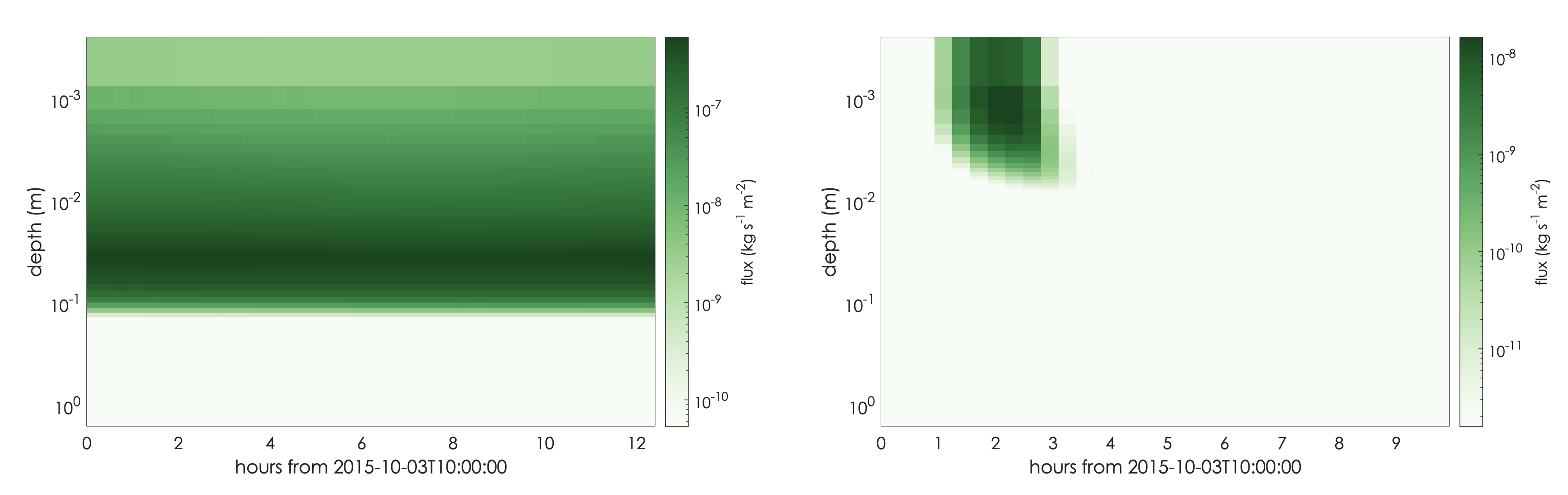}
\caption{Flux rate of water vapour during one rotation of the nucleus starting from 2015-10-03T10:00:00 at different depths of (a) facet S and (b) facet N}
\label{fig:flux}
\end{figure}

As a consequence, sublimation of volatile ices also contrasts between the two facets. Figure \ref{fig:flux} shows the intensity of water ice sublimation on the day of boulder migration. The mean flux rate of water vapour at facet S was $\sim3.2\times 10^{-5}\text{ kg m}^{-2}\text{s}^{-1} $, three orders of magnitude higher than that at facet N, meaning that gas emission was occurring mainly, if not only on the southern side of the boulder. With uneven gas emission over the boulder, ``rocket force" could be at work, as observed on decimeter-sized boulders in the near-nucleus coma \citep{2016MNRAS.462S..78A}. According to the equation (2) in \citet{2016MNRAS.462S..78A}, the above flux rate could yield an acceleration of $\sim2\times10^{-5}\text{ m s}^{-2}$, one order of magnitude smaller than the local gravitational acceleration of $\sim2\times10^{-4}\text{ m s}^{-2}$.

It is interesting to notice that the bulk of sublimation at facet S comes from a larger depth of about five centimeters (Figure \ref{fig:flux}a). This is due to the persistent illumination on the southern side of the boulder that desiccated the top few centimeters of subsurface. As the heat wave penetrates to deeper interior of the boulder, it could reach reservoirs of more volatile ices, such as carbon dioxide, causing more intensive sublimation activity. Previous works demonstrated the triggering of super-volatile sublimation could be the mechanism driving outbursts on 67P \citep{2016A&A...593A..76S,2024MNRAS.529.2763M}.

\section{Discussion} \label{sec:discussion}
The wealth of imaging data acquired by Rosetta enabled us to constrain the time of this boulder migration event to within $\sim$14 hours. However, given the large distance between the spacecraft and the nucleus at the time, direct observation of the boulder's motion is lacking, leaving this extraordinary cometary geological event to be enigmatic. Nevertheless, our new analysis and modelling results shed light on the possible scenarios for the boulder's mobilisation, as discussed below:

\subsection{Scenario A: Ejection by outburst}
As proposed in previous works, direct outburst stays the most straightforward scenario responsible for the displacement of the boulder \citep{el2017surface}, in which outgassing at the boulder's original location transported it to its new location via gas drag force.

To lift a 30-m boulder, taking particle size and boulder shape into consideration, \cite{el2017surface} estimated that the minimum gas production rate required is $25$ kg$\cdot$s$^{-1}$ when the driving agent was H$_{2}$O. This amounts to around 5.5\% of the global production rate of 67P at the time of the migration event, which is $\sim 450 \text{ kg}\cdot$s$^{-1}$  \citep{2016MNRAS.462S.491H}. 
An outburst at such scale would likely to release large quantity of dust particles, that could transiently increase the brightness of the coma significantly. For comparison, the gas production rate of the outburst on 2016 February 19 is estimated to be $10$ kg$\cdot$s$^{-1}$ \citep{2016MNRAS.462S.220G}, and it caused an increase in the brightness of 67P's coma by two orders of magnitude (Figure 4 in \cite{2016MNRAS.462S.220G}). 
Since the brightness surge usually lasts for less than half an hour, and given the sparse observations at the time, we are not able to conclude whether such an outburst did occur within the 14 hour window of the migration event.
Outbursts are also commonly accompanied by surface changes, often revealing more pristine subsurface with higher ice content. The best example might be the collapse of Aswan cliff that revealed fresh nucleus material with albedo six times that of the bulk value \citep{pajola2017pristine}. Surface changes also occur following the transition of water ice from the amorphous to the crystalline state. The outburst on 2016 July 3, with an estimated gas production rate of $(18.4 \pm 10.6)$ kg $\cdot$ s$^{-1}$ affected an area of $\sim 10$ m radius, creating an ice patch of $15 \times 5 \text{ m}^2$ \citep{agarwal2017evidence}. In the case of Khonsu boulder, the `after' observations do not show prominent increase in albedo at the original location of the boulder (Figure \ref{fig:boulder_migrate} c0 \& d0). However, the resolution of $\sim 15$ m hinders quantitative analysis on the surface change immediately after the boulder migration, if there was any.
While numerous surface changes were found near the boulder, as described in section \ref{subsec:geomorphology}, it is not clear if these changes are directly related to the boulder migration event.

\subsection{Scenario B: Destabilisation by nearby activities}

Activities in neighbouring areas could destabilise the boulder and trigger its downslope movement either by eroding the slope it resided on \citep{el2017surface}, or by seismic shaking \citep{hasselmann2019pronounced}. 
Rosetta has observed many significant surface changes in Khonsu region after comet perihelion, including formation of cracks and pits, cliff collapses, and the boulder migrated into the northern high bank \citep{hasselmann2019pronounced}.
As the cohesion of surface materials on comet is considerably weaker than that on Earth, a small activity can induce perturbation that has a relatively large impact in proximity, as the theoretical calculations of P and S wave velocities are comparable to the escape velocity \citep{vincent2016summer}.
Indeed, the night-time mini outbursts we found on both the day of the migration (Oct. 03) might perform as the source of destabilisation.
It has been noticed that the original location of the boulder was close to the source of a massive outburst that occurred on 2015 December 13. This outburst revealed an ice-rich patch on the surface and had a similar behavior as another outburst that occurred on  2016 July 3 \citep{hasselmann2019pronounced}.
Seismic shaking has been observed on several asteroids, and was suggested as the cause for regolith transport \citep{2020Icar..34713811R,miyamoto_regolith_2007}. Impact-induced vibrations can play an important role in the evolution of regolith on irregular-shape small bodies, for the the seismic acceleration caused is comparable to surface gravity and responsible for gravel destabilization (Figure 4 of \cite{miyamoto_regolith_2007}). 
Yet the propagation of seismic wave significantly depends on the mechanical properties of the body. Bodies like Ryugu shows extremely low seismic efficiency as was found with the SCI impact by Hayabusa2 \citep{2021JGRE..12606594N,2019Icar..331..179M}. In fact, 67P shows numerous  `boulders-on-boulders' features similar to those on Ryugu, that might imply a similarly low seismic efficiency (Figure \ref{fig:boulder_perch}).

\subsection{Scenario C: boulder's own activity}

Here we propose a third possible scenario responsible for the migration, which involves the boulder's own activity.
Noticeable boulders ($\geq$ 7 meter) on 67P are presumed to be ice-rich and can be an inventory of volatiles \citep{pommerol_osiris_2015,pajola2015size}, just like the comet nucleus  \citep{rubin_volatiles_2023}. The migrating boulder in Khonsu, with such a distinctive 30-meter scale, may preserve different volatiles at different depths.
As the thermophysical analysis shows, the boulder was subjected to heterogeneous insolation conditions that resulted in asymmetric energy accumulation and thermal history between its southern and northern sides.
Profiles of orbital temperature and water flux variation show that the mobilisation event coincided with peaking of temperatures at the depth of 10 centimeter on the boulder's southern side, reaching over 200 K.
Given the likely existence of highly-volatile ices at that depth, the southern side of the boulder could experience explosive outgassing events. Meanwhile, the northern side was undergoing its  `winter', with temperatures staying below 120 K at all depths.
This dichotomy in activity would lead to a contrast in outgassing flux on the two sides, exerting a net propulsion towards the north.
In fact, the boulder performed much as a miniature nucleus. Globally, 67P's southern hemisphere experienced active summer during perihelion passage, the intense solar heating led to the morphological and topographic diversity between the two hemisphere and triggered peak dust activity, introducing mass transport across nucleus from south to north. 
In this scenario, the source of the mini-outburst on October 3 (Figure \ref{fig:act_timeline}a) could be the boulder itself. It would serve as an unexhausted source for the mini outburst on October 04 as well (Figure \ref{fig:act_timeline}$ b1 \sim b4$) . Or, the dust activity on October 4 might be sustained by newly exposed icy areas on the boulder due to its displacement.

\subsection{Comparison with the other Khonsu ``jumping boulder"}
\citet{hasselmann2019pronounced} found in the northern high bank of Khonsu another noticeable ``jumping boulder" with even larger dimension of $\sim$50 m. They argued that it was likely initially ejected by an outburst somewhere outside Khonsu region, travelled over the nucleus before falling back onto the surface. It possibly performed a gentle impact on the nearby cliff before touchdown while maintaining its intact structure. 

Similar to the boulder discussed in this work, the northern jumping boulder’s inferred movement is temporally aligned with a series of activities that could have been linked to its movement or interaction with the cliff. 
However, the relationship between the boulder migration event discussed in this work and surrounding dust activities is less direct. The only concrete observation during the migration process is linked to a night-time mini-outburst, the source of which remains unclear. It is possible that the activity is from the boulder itself if the motion was triggered by its own activity, or the dust activity could be from its new location triggered by the boulder's impact. The dust emanation observed in the following days at the beginning of October could be activities from the freshly exposed icy surfaces of the migrated boulder or its surroundings.

Given the similar observational evidence, it is possible that the two Khonsu ``jumping boulders" could share a similar movement triggering mechanism. Both boulders could be involved in a scenario of direct ejection by outburst, with a high gas production rate comparable to that reported in \cite{agarwal2017evidence}, or they could have been triggered and propelled by their own activity. However, as we do not know the original position of the other boulder in the high bank, it is difficult to compare quantitatively the two events. 

\section{Conclusion and outlook}
- We revisited the relocation of a $\sim$30m boulder in the Khonsu region of comet 67P. With a $\sim$140m displacement, this is the most prominent boulder migration event observed on 67P.

- Utilizing a high-resolution shape model, we derived a detailed timeline for the event, narrowing the occurrence of migration to within 14 hours on 2015 October 3, roughly two months after 67P's perihelion.

- Night-time mini-outbursts are found, both during the migration period, and in the following day, near the new location of the boulder.

- The boulder and its surroundings displayed a substantial morphological changes based on comparisons of pre- and post-perihelion observations. 

- Thermophysical modelling shows that before the boulder’s migration, cumulative insolation on southern and northern sides of the boulder had different patterns. Analysis of the thermal history of the boulder identified a significant north-south dichotomy in the variation of temperature and flux rate.
The displacement co-occurred with maximum temperature on the southern side 10 centimeters below the surface.

- New findings neither confirm nor invalidate hypotheses for the boulder's migration in previous works. However, in addition to direct outburst and seismic vibration, we propose a third mechanism for the triggering of the boulder movement, which is by the propulsion due to intense outgassing on its southern side. However, the actual scenario might be a combination of several triggers.

- Displacement of large boulders on cometary surfaces presents a unique opportunity for probing cometary activity as well as a nucleus' physical properties. Boulder-sized debris ranging from decimeter to meter sizes were discovered being ejected and falling back,  \citep{shi2024diurnal,2016MNRAS.462S..78A}, potentially driven by the sublimation of more volatile ices, such as carbon dioxide. For the displacement of even larger chunks like the one discussed here, mechanisms could be diverse and need to be more thoroughly explored, particularly regarding their interactions with local topography and dust activity. To this end, we plan to reconstruct time-varying topography of local Khonsu region  to quantitatively depict changes of the boulder and its surroundings. This may also help in reconstructing a possible migration scenario using numerical methods.

\begin{acknowledgments}
The authors thank an anonymous referee for insightful comments that helped improve the manuscript significantly. This work is supported by the National Natural Science Foundation of China (No. 12233003). MRELM acknowledges funding from the KU internal grant (8474000336-KU-SPSG). XS thanks members of the ISSI International Team ``Closing The Gap Between Ground Based And In-Situ Observations Of Cometary Dust Activity: Investigating Comet 67P To Gain A Deeper Understanding Of Other Comets" for inspirational discussions that helped to improve this study. OSIRIS was built by a consortium led by the Max-Planck-Institut f{\"u}r Sonnensystemforschung, G{\"o}ttingen, Germany, in collaboration with CISAS, University of Padova, Italy, the Laboratoire d'Astrophysique de Marseille, France, the Instituto de Astrof{\'i}sica de Andalucia, CSIC, Granada, Spain, the Scientific Support Office of the European Space Agency, Noordwijk, The Netherlands, the Instituto Nacional de T{\'e}cnica Aeroespacial, Madrid, Spain, the Universidad Polit{\'e}chnica de Madrid, Spain, the Department of Physics and Astronomy of Uppsala University, Sweden, and the Institut f{\"u}r Datentechnik und Kommunikationsnetze der Technischen Universit{\"a}t Braunschweig, Germany. The support of the national funding agencies of Germany (DLR), France (CNES), Italy (ASI), Spain (MEC), Sweden (SNSB), and the ESA Technical Directorate is gratefully acknowledged.

\end{acknowledgments}

\bibliography{2023_boulder_migration}{}
\bibliographystyle{aasjournal}

\appendix

\begin{figure}[h]
\centering
\includegraphics[width=0.8\textwidth]{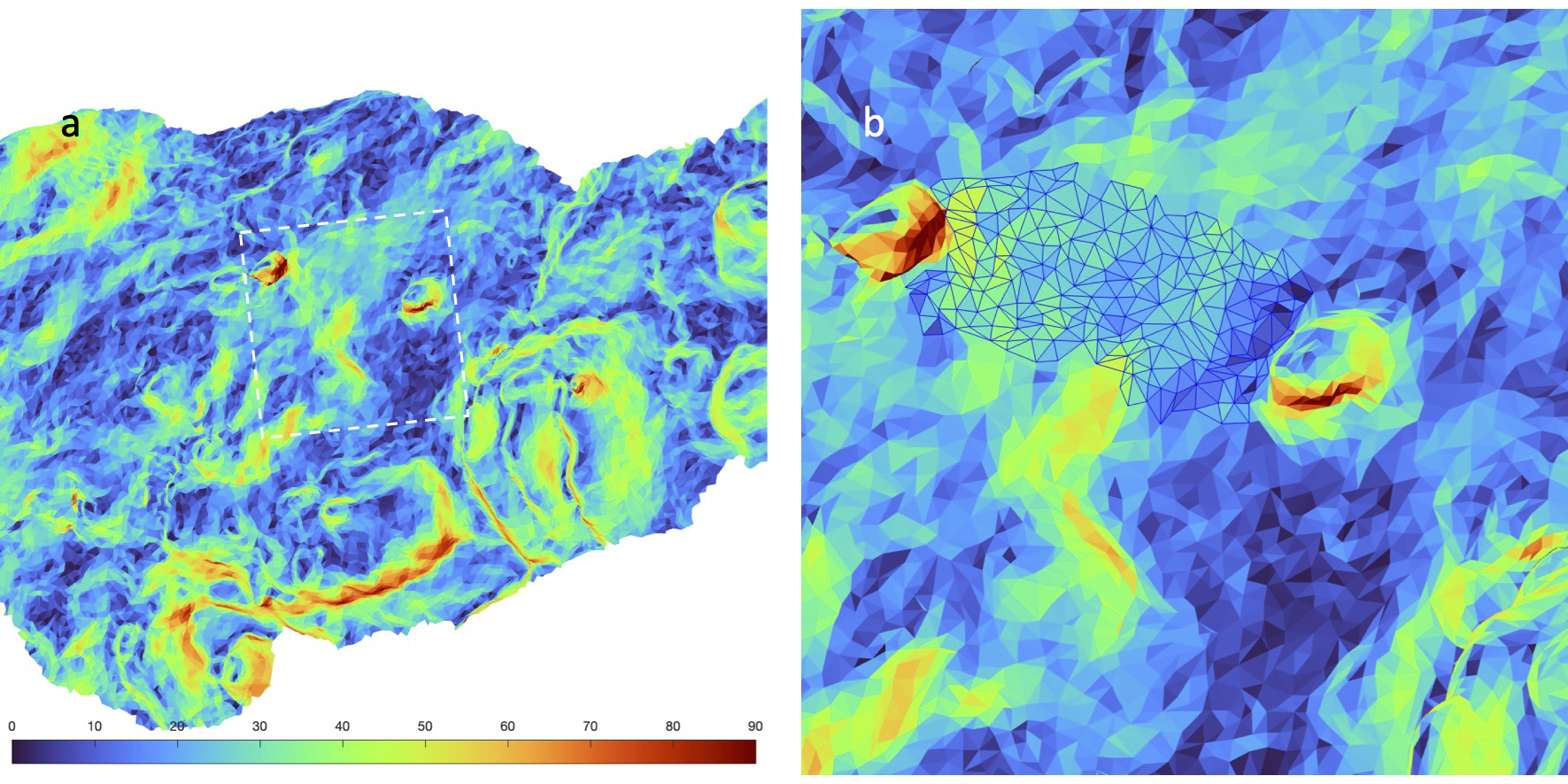}
\caption{Effective surface slope in the areas surrounding the migrating boulder. The effective surface slope was defined as the angle between the negative surface normal direction and the vector of surface acceleration that consists of both gravitational and rotational accelerations. (a) Surface slope in the Khonsu region. (b) The slope between the boulder's original location and new location.}
\label{fig:ks_slp}
\end{figure}

\begin{figure}[h]
\centering
\includegraphics[width=0.6\textwidth]{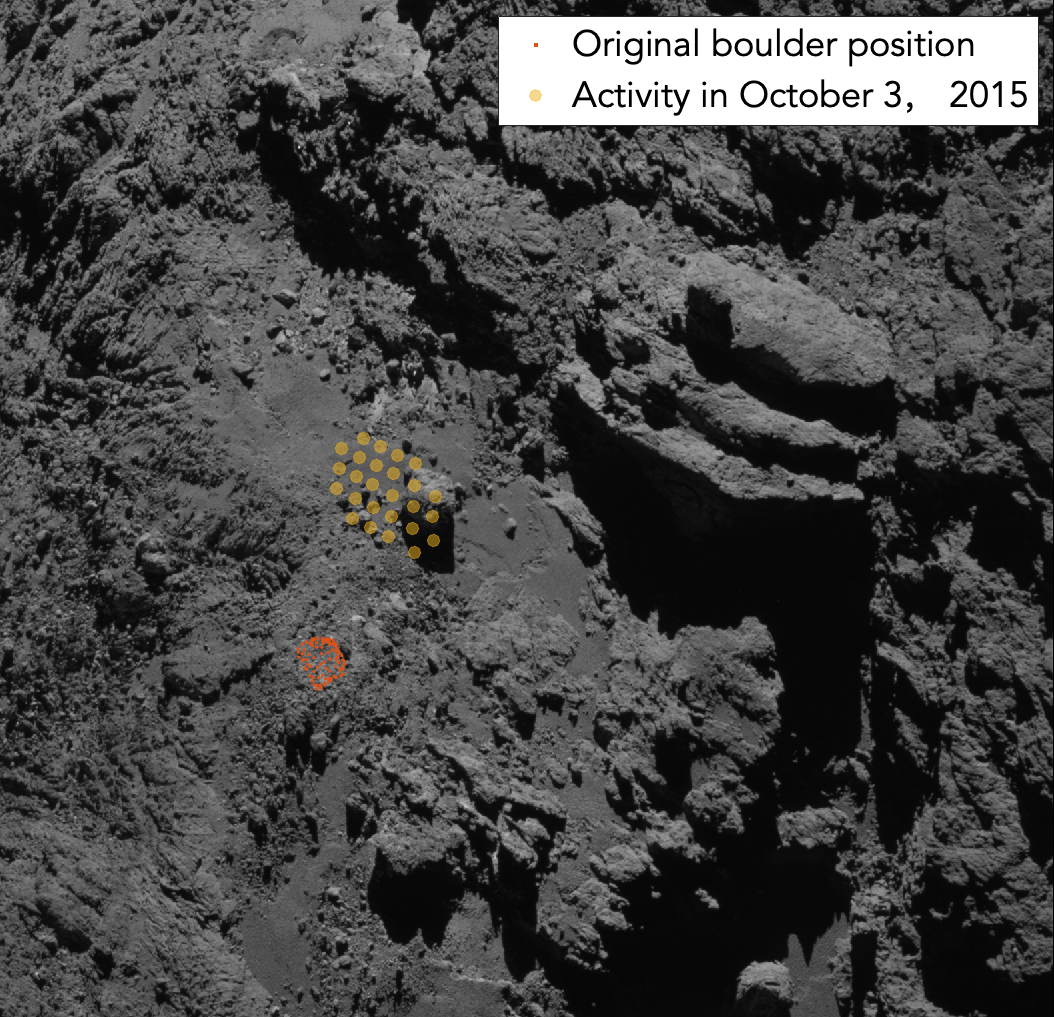}
\caption{Original boulder position and mini-outburst projected onto the NAC image of 2016 June 11, UTC 05:11. Position of pre-migration boulder is represented in red dots created from vertices of its shape model, while the orange markers show the reprojection of night-time dust emission in 2015 October 3, UTC 19:30.}
\label{fig:mark_oct}
\end{figure}

\begin{figure}[h]
\centering
\includegraphics[width=\textwidth]{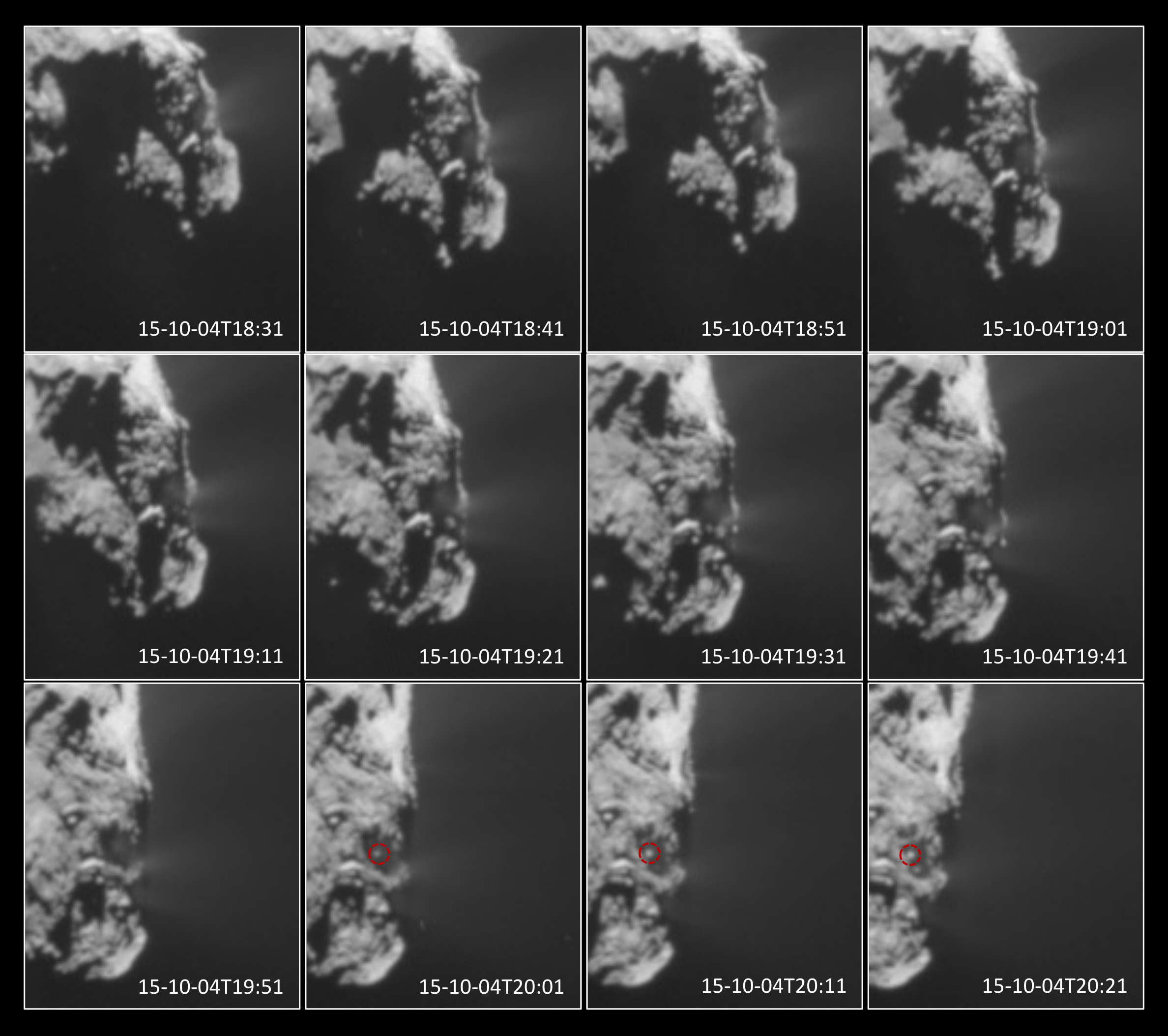}
\caption{The sequence MTP021\_STP076\_OUTBURST\_003, observations for the prolonged activity after migration event in October 4.}
\label{fig:act_oct}
\end{figure}

\begin{figure}[h]
\centering
\includegraphics[width=\textwidth]{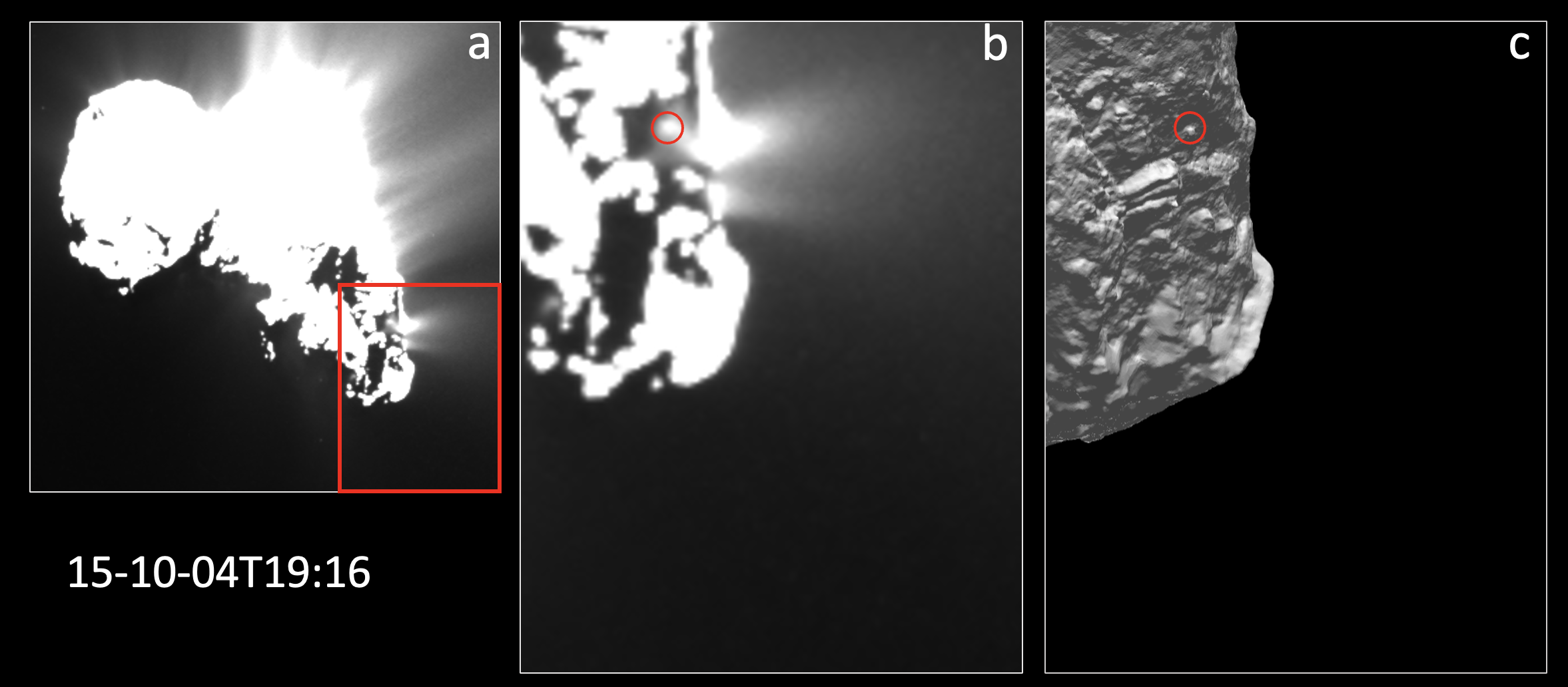}
\caption{Mini-outburst on 2015 October 4, and its possible correlation with the boulder. (a) Actual observation taken at UTC 19:16 with long exposure time to reveal dust activity and possible outbursts on 67P. (b) Enlarged view of part of the image in the red frame in (a), showing a mini-outburst from the unilluminated smooth area near the ``pancake" feature. Red circle indicates the possible source of the mini-outburst being its brightest spot. (c) Synthetic image showing the same part as (b). Artificial lighting is applied to reveal all surface features. The red circle is at the same position as in (b), which coincides with the position of the boulder after migration.}
\label{fig:minioutburst}
\end{figure}

\begin{figure}[h]
\centering
\includegraphics[width=0.5\textwidth]{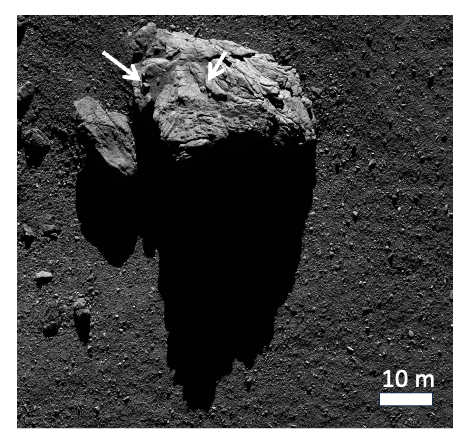}
\caption{OSIRIS NAC image taken on 2016 September 20, showing a $\sim$60 m boulder in Hatmehit region. On top of this boulder, several smaller meter-scale boulders are observed, and two of which are indicated by arrows.}
\label{fig:boulder_perch}
\end{figure}

\end{document}